\documentclass[reprint]{revtex4-2}
\pdfoutput=1
\usepackage{graphicx}
\usepackage{dcolumn}
\usepackage{bm}
\usepackage{color}
\usepackage{amsmath}
\usepackage{amssymb}
\usepackage{comment}
\usepackage[utf8]{inputenc}
\usepackage{textcomp}
\usepackage{braket}
\usepackage{soul}
\usepackage[hidelinks]{hyperref}
\usepackage{xr}
\usepackage[symbol]{footmisc}

\usepackage{subfiles}
\usepackage{blindtext}
\usepackage{amsmath,amsfonts,amsthm,bm} 
\usepackage{todonotes}
 
\usepackage{float}
\newfloat{suppfig}{tbh}{losf}
\floatname{suppfig}{\textcolor{blue}{Supplementary} Figure}
\usepackage[a4paper, total={6.9in, 10.02in}]{geometry}
\usepackage{float}
\usepackage{orcidlink}
\usepackage{lipsum}
\usepackage{booktabs}
\usepackage{multirow}
\usepackage{makecell}
\usepackage{tabularx}
\usepackage{needspace}
\usepackage[normalem]{ulem}
\newcommand\emailx[1]{%
\move@AF%
\def\@affil{{\normalfont\,#1\strut}{}}%
}%

\makeatletter
\newcommand{\fmarki}{*}
\newcommand{\fmarkii}{\ensuremath{\dagger}}
\def\@fnsymbol#1{{\ifcase#1\or \fmarki\or \fmarkii\else\@ctrerr\fi}}
\makeatother
\renewcommand{\fmarki}{}
\renewcommand{\fmarkii}{}

\begin{document}

\title{Confocal imaging from biphoton correlations}

\thanks{\parbox{\linewidth}{\raggedright * Corresponding author: \href{mailto:miles.padgett@glasgow.ac.uk}{miles.padgett@glasgow.ac.uk}}}

\author{Euan Millar~\orcidlink{0009-0009-2078-6067}}
\author{Emma Pearce~\orcidlink{0000-0002-5407-2187}}
\author{Daniele Faccio~\orcidlink{0000-0001-8397-334X}}
\author{Miles J. Padgett~\orcidlink{0000-0001-6643-0618}*}

\affiliation{School of Physics and Astronomy, University of Glasgow, Glasgow G12 8QQ, UK}


\begin{abstract}
\noindent Confocal microscopy provides optical sectioning for three-dimensional imaging but conventionally relies on point-by-point scanning and physical pinholes, limiting imaging speed. Here, we demonstrate that confocal sectioning can instead arise directly from quantum measurement, using spatial correlations in entangled biphoton states. Spatially resolved coincidence measurements in a widefield imaging system suppress contributions from out-of-focus planes, producing optical sectioning in parallel across the whole field of view without physical pinholes or mechanical scanning. In addition, this quantum approach obtains an axial response that, as a result of position-momentum entanglement, is $\sqrt{2}$-times narrower than that of classical confocal imaging. Our work establishes biphoton correlations as a mechanism for improved confocal imaging and enables parallel, pinhole-free optical sectioning without mechanical scanning.
\end{abstract}


\maketitle


\clearpage

\noindent Confocal microscopy, the canonical approach to optically sectioned imaging, combines pinhole-based spatial filtering in the illumination and detection pathways to achieve three-dimensional imaging capabilities~\cite{Sheppard1988}. By rejecting out-of-focus light, confocal imaging systems provide high-resolution depth discrimination and have become central to biological and materials imaging. However, this sectioning mechanism is inherently tied to scanning-based implementations, which limit imaging speed. Complementary approaches, including structured illumination~\cite{Saxena15, Li2020} and light-sheet microscopy~\cite{LSM, Andilla2018}, provide routes to optical sectioning through engineered illumination patterns or specialized setups, but are limited to distinct imaging regimes and length scales~\cite{Elliott2020}. In all of these approaches, depth selectivity is achieved through spatial filtering and illumination shaping. This raises a more fundamental question: can optical sectioning emerge from the measurement process alone without modifying the optical field itself? 

Quantum correlations offer an additional resource for imaging. Ever since the original analysis of nonlocal correlations by Einstein, Podolsky, and Rosen~\cite{Einstein1935}, entanglement has been understood to impose constraints on joint measurement outcomes that cannot be reproduced classically. In light, such correlations are routinely generated through spontaneous parametric down-conversion (SPDC) and have given rise to quantum imaging techniques such as ghost imaging~\cite{Shih1995}. Subsequent work has also demonstrated improvements in signal-to-noise ratio~\cite{Brida2010} and transverse resolution~\cite{Tsang2009, Toninelli2019}, establishing entangled photon pairs as a powerful resource for imaging. However, existing quantum imaging approaches have primarily exploited correlations to modify transverse resolution, improve sensitivity, or to access phase information~\cite{Padgett2017, Chesterking2019, Gregory2025, Defienne2019, Kaur2021, Devrari2025}. More recent studies have demonstrated enhanced axial resolution~\cite{Tenne2019} and reshaped axial transfer functions to extend the depth of field~\cite{Zhang2024}; however, in these cases, quantum correlations modify the performance of an existing scanning imaging modality rather than generate the sectioning modality itself. 

Here we show that confocal sectioning can arise entirely by quantum measurement. One way of understanding these quantum correlations is through the advanced wave picture of Klyshko~\cite{Klyshko1988}. In this interpretation, the detection process can be viewed as a back-propagation of the detected photons through the optical system. The light in the optical system is therefore treated as originating from the detected position of the idler (signal) photon, propagates through the optical system, is reflected from the SPDC source, and recorder at the detected position of the signal (idler). The recorded signal from this is proportional to the coincidence count. The spatial correlations impose an effective pinhole, selecting only photon pairs originating from the in-focus plane. This produces a confocal effect without any scanning pinholes. 

In a transmission imaging experiment, we observe a narrowing of the axial point spread function (PSF) across the full field of view. Despite the absence of any pinholes, the measured axial response shows a narrowing compared to standard widefield imaging that is not only consistent with classical confocal detection but also shows a further narrowing that can be explained as a result of position-momentum entanglement. This progression---from widefield to confocal to biphoton imaging---corresponds to axial PSF narrowing factors of $1:\sqrt{2}:2$, respectively.

By identifying entangled photon (biphoton) correlations as a mechanism for confocal imaging, this work reframes confocal imaging as a natural consequence of biphoton detection, while nonseparability provides an additional phase-based enhancement of axial localization. 

\begin{figure*}
    \centering
    \includegraphics[width=\linewidth]{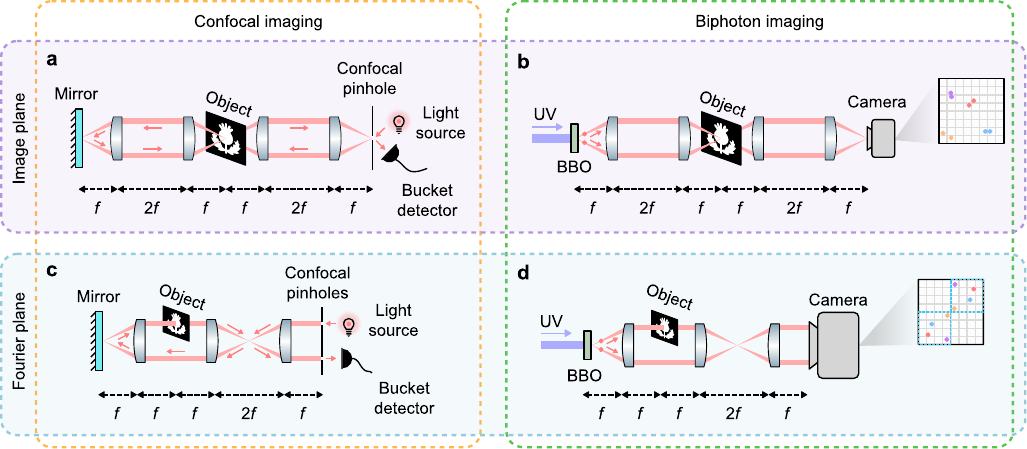}
    \caption{\textbf{Image- and Fourier-plane configurations for classical confocal and biphoton imaging.} (\textbf{a}, \textbf{b})~Image-plane implementations for confocal (\textbf{a}) and biphoton (\textbf{b}) imaging, with entangled photon pairs indicated by matched colors. \mbox{(\textbf{c}, \textbf{d})}~Corresponding Fourier-plane implementations for confocal (\textbf{c}) and biphoton (\textbf{d}) imaging, with the same color coding of correlated events. In both regimes, the optical layouts are geometrically analogous. Within a retrodictive framework, the nonlinear crystal (BBO) acts as an effective mirror, establishing a direct correspondence between biphoton coincidence detection and classical confocal spatial filtering. Distances are specified in units of the focal length $f$ of the imaging optics.}
    \label{fig:concept}
\end{figure*} 


\vspace{0.5 cm}
\noindent\textbf{\large{Confocal imaging as a consequence of biphoton correlations}}

\noindent The defining feature of confocal imaging is the multiplicative combination of the illumination and detection responses. At transverse coordinate $\mathbf{r}$, the detected signal in a classical confocal microscope is given by \\
\begin{equation}
    I_\mathrm{{conf}}(\mathbf{r}) \propto I_{\mathrm{ill}}(\mathbf{r}) \cdot I_{\mathrm{det}}(\mathbf{r}).
\end{equation} \\
For identical illumination and detection pathways, with $h(\mathbf{r})$ denoting the corresponding field PSF, this reduces to
\begin{equation}
    I_\mathrm{{conf}}(\mathbf{r}) \propto \left|h\right|^4 = \left|I_{\mathrm{wf}}(\mathbf{r})\right|^2,
    \label{eq:conf}
\end{equation} 
where $I_{\mathrm{wf}} = |h|^2$ is the classical widefield intensity PSF (see Supplementary Information for full derivation). This increased selectivity leads to optical sectioning, with the theoretical axial resolution improving from \mbox{$D_z = 2n\lambda/\mathrm{NA}^2$} in widefield microscopy to $D_z \simeq \sqrt{2}n\lambda/\mathrm{NA}^2$ for a confocal microscope assuming a Gaussian profile. 
Optical sectioning therefore arises from the spatial selectivity imposed by the combined illumination and detection pathways of the imaging system~\cite{Pawley2006}. 

The same spatial filtering can be understood from the perspective of correlated photon detection. Inspired by the Klyshko interpretation~\cite{Klyshko1988}, the geometric correspondence between classical confocal detection and biphoton coincidence imaging is illustrated in Fig.~\ref{fig:concept}. We show here, in both image- (Fig.~\ref{fig:concept}a, b) and Fourier-plane (Fig.~\ref{fig:concept}c, d) representations, that the spatial filtering normally imposed by the confocal pinhole can instead be implemented via spatially correlated detection of the photon pairs. 

For degenerate photons generated by SPDC, the biphoton state after propagation over a distance $z$ can be written in transverse-momentum space as \\
\begin{equation}
    \ket{\Psi(z)} =
    \iint d^2\mathbf{q}_s \ d^2\mathbf{q}_i \
    \Phi(\mathbf{q}_s,\mathbf{q}_i)
    e^{i z k_z(\mathbf{q}_s)}
    e^{i z k_z(\mathbf{q}_i)}
    \ket{\mathbf{q}_s,\mathbf{q}_i},
    \label{eq3}
\end{equation} \\
where $\Phi(\mathbf{q}_s,\mathbf{q}_i)$ is the initial biphoton amplitude at the crystal output $z=0$ and $\mathbf{q}_{s,i}$ denotes the transverse momentum vector of the signal and idler photons. The two propagation factors $e^{i z k_z(\mathbf{q}_{s,i})}$ describe the phase accumulated independently by the signal and idler photons, respectively~\cite{Hong1985, Monken1998}. 
%
%
%
%
%

The coincidence measurement probes the joint spatial detection amplitude, \\
\begin{equation}
    \psi^{(2)}(z) = \braket{\mathbf{r}_s, \mathbf{r}_i | \Psi(z)},
\end{equation} \\
with a coincidence probability given by
\begin{equation}
    G^{(2)}(\mathbf{r}_s, \mathbf{r}_i;z) \propto \left|\psi^{(2)}(z)\right|^2.
\end{equation} 
%
%

For a separable state, $\Phi(\mathbf{q}_s, \mathbf{q}_i) = \phi(\mathbf{q}_s)\phi(\mathbf{q}_i)$ and the joint detection amplitude factorizes into the product of the two single-photon amplitudes. For identical optical responses, each single-photon detection amplitude can be denoted as the field PSF $h(\mathbf{r};z)$, introduced above. Therefore, when both photons are detected at the same transverse position, $\mathbf{r}_s=-\mathbf{r}_i = \mathbf{r}$, this gives \\
\begin{equation}\label{g2_sep}
    G^{(2)}_{\mathrm{sep}}(\mathbf{r}, \mathbf{r};z) \propto |h(\mathbf{r};z)|^4 = I^2_{\mathrm{wf}}(\mathbf{r};z).
\end{equation} \\
Thus, the familiar fourth-power response of confocal imaging can arise from coincident detection of two separable photons and does not, in itself, require entanglement (see Supplementary Information for the full derivation).

However, when the biphoton state is non-separable, the two propagation phases contribute coherently to a single joint two-photon detection amplitude. Photons generated by SPDC are strongly momentum anticorrelated, such that \mbox{$\mathbf{q}_s \approx -\mathbf{q}_i$}. In a symmetric imaging system, \mbox{$k_z(\mathbf{q}_s)$ = $k_z(-\mathbf{q}_i)$ = $k_z(\mathbf{q}_i)$}, and hence the joint propagation phase is \\
\begin{equation}
    e^{izk_z(\mathbf{q}_s)}e^{izk_z(\mathbf{q}_i)} = e^{i2zk_z(\mathbf{q})}.
\end{equation} \\
Thus, the axial evolution of the biphoton field is governed by the sum of the longitudinal wavevectors, \mbox{$k_{z,s} + k_{z,i}$}, which is twice that of the single-photon value. 

Hence, the two photons acquire a combined longitudinal phase corresponding to twice the axial propagation of a single photon. The resulting two-photon detection amplitude therefore has the same axial dependence as the single-photon field PSF evaluated at twice the displacement (see Supplementary Information for the full derivation). Taking the modulus squared with $\mathbf{r}_s=\mathbf{r}_i = \mathbf{r}$ then gives \\
\begin{equation}\label{g2_ent}
    G^{(2)}_{\mathrm{ent}}(\mathbf{r}, \mathbf{r};z) \propto |h(\mathbf{r};2z)|^2 = I_{\mathrm{wf}}(\mathbf{r};2z).
\end{equation} 
%

This result indicates that non-separable quantum states behave in a fundamentally different manner compared to separable  momentum-position-correlated states, Eq.~\eqref{g2_sep} or standard scanning confocal imaging with uncorrelated states, Eq.~\eqref{eq:conf}.  
In standard confocal imaging, spatial selectivity arises from the multiplication of illumination and detection intensities, yielding $|h|^4$. This same dependence is recovered in Eq.~\eqref{g2_sep} from biphoton position-momentum correlations, which is ultimately the origin for a similar $|h|^4$ dependence. In the entangled-photon case, the two-photon amplitudes are instead combined coherently before taking the modulus squared, leading to a markedly different $|h(2z)|^2$ dependence. The resulting axial compression is therefore a consequence of the quantum state and its propagation phase, rather than of the multiplicative intensity filtering that underlies classical confocal imaging. When considering a Gaussian beam, the scaling of the quantum confocal response yields a $2\times$ narrowing of the axial localization relative to widefield imaging and a $\sqrt{2}$ improvement over classical confocal imaging (full derivation in the Supplementary Information). Equivalently, under this Gaussian assumption, the quantum case can be thought of as an effective $|h|^8$ or $I_{\mathrm{wf}}^4$ scaling to provide a comparison with the classical case.

\begin{figure}
    \centering
    \includegraphics[width=\linewidth]{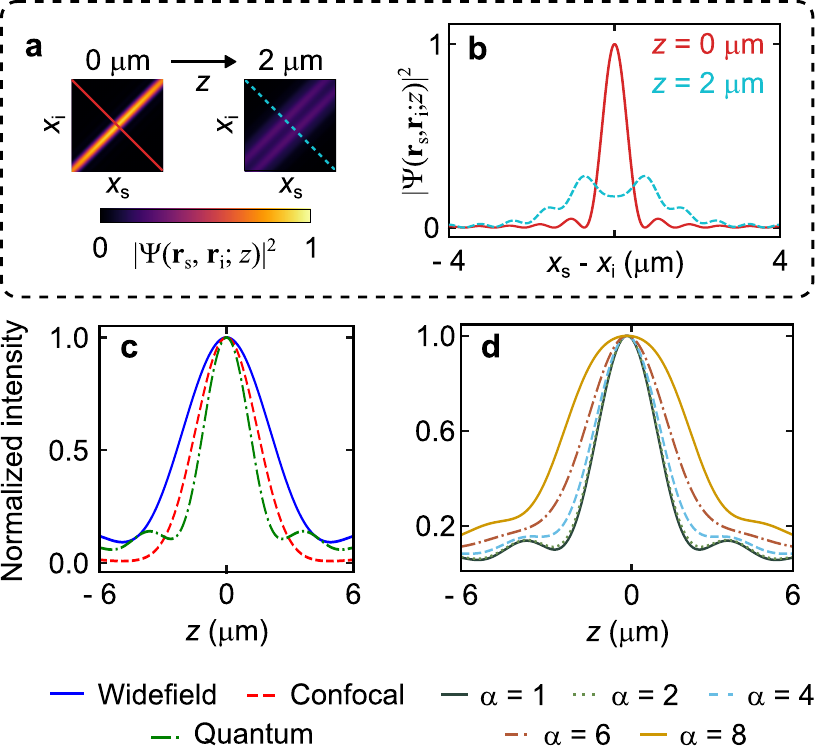}
    \caption{\textbf{Correlation-induced confocal sectioning effect.} (\textbf{a})~Simulated biphoton joint probability distributions at the focal plane and $z=2$ \textmu m defocused. (\textbf{b})~Corresponding cross-diagonal profiles. (\textbf{c})~Normalized axial PSFs for widefield, confocal, and biphoton imaging, showing enhanced optical sectioning arising from spatially correlated detection. (\textbf{d})~Dependence of the biphoton PSF on the transverse correlation strength, modified by tuning the phase-matching function by a factor $\alpha$. 
    }
    \label{fig:psf}
\end{figure}

Figure~\ref{fig:psf} provides a graphical representation of the expected behavior of quantum confocal imaging. Figure~\ref{fig:psf}a shows simulated biphoton joint probability distributions at the focal plane and under axial defocus. At the focus, the probability distribution is strongly concentrated along the anti-diagonal, whereas axial defocus broadens and redistributes the correlated signal. Examining cross-diagonal profiles reveals a pronounced reduction in coincidence counts with increasing defocus (Fig.~\ref{fig:psf}b). This dependence of the counts on the axial position provides the basis for the optical sectioning capability of quantum confocal imaging. As shown in Fig.~\ref{fig:psf}c, the combination of spatial correlations and addition of the longitudinal wavevectors produces an axial PSF that is narrower than those obtained with both widefield and conventional confocal imaging, resulting in enhanced optical sectioning. 

The axial confinement is determined by the transverse correlation bandwidth, subject to the available spatial-frequency range imposed by the pupil. To illustrate this dependence, we vary the transverse phase-matching bandwidth by introducing a scaling factor $\alpha$ in the phase-matching function in our simulation. Increasing $\alpha$ narrows the transverse phase-matching bandwidth and produces a correspondingly broader axial response (Fig.~\ref{fig:psf}d). Conversely, increasing the transverse phase-matching bandwidth enhances the range of longitudinal wavevectors contributing to the biphoton field and thereby strengthens axial confinement.

\begin{figure*}
    \centering
    \includegraphics[width=\linewidth]{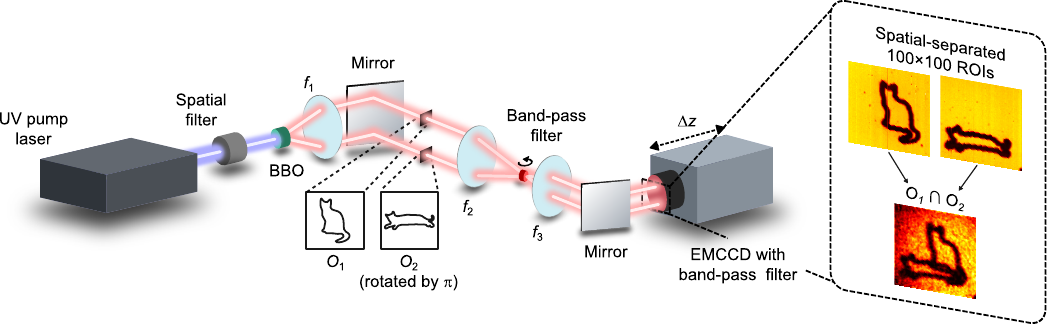}
    \caption{\textbf{Far-field biphoton imaging setup.} A 355~nm pump laser is spatially filtered and directed onto a BBO crystal to generate spatially entangled photon pairs via type-II SPDC. The far-field of the down-converted biphoton is relayed ($f_1$) through a demagnifying telescope ($f_2$ and $f_3$) and imaged onto an EMCCD camera, where two regions of interest, $O_1$ and $O_2$, are selected. A $\pi$ rotation is applied to $O_2$ to convert momentum anti-correlations into spatial correlations at the detector. Band-pass and long-pass filters select near-degenerate photon pairs, and the camera is scanned along the optical axis to record images at different focal planes. Insets show the two objects as projected in the far-field plane and the resulting coincidence measurement ($O_1 \bigcap O_2$).}
    \label{fig:setup}
\end{figure*}


\needspace{2\baselineskip}
\vspace{0.5 cm}
\noindent\textbf{\large{Experimental realization of quantum confocal imaging}}

\noindent To demonstrate the quantum confocal effect experimentally, we implemented a far-field, biphoton imaging system (Fig.~\ref{fig:setup}, see Methods for full details). Type-II SPDC in a 3-mm long $\beta$-barium-borate (BBO) crystal generates momentum-anticorrelated photon pairs, which are spatially separated in the Fourier plane and detected in separate regions of interest (ROIs) on an electron-multiplying charge-coupled device (EMCCD) array. 

Opaque objects are placed in the signal and idler paths ($O_1$ and $O_2$), such that coincidence images encode features from both paths simultaneously, providing a direct visualization of the joint nature of the two-photon measurement rather than independent imaging in either arm (as shown in the inset in Fig.~\ref{fig:setup}). The use of two objects here is solely for illustrative purposes; the same imaging principle applies in the standard configuration with a single object in one path and free propagation in the other. Since the signal and idler photons are anticorrelated, to correlate the detected events from each region the idler image is rotated by $\pi$, converting transverse momentum anticorrelations into spatial correlations in a reassigned coordinate frame. Coincident photon pairs are then mapped onto a common bisector coordinate (see Supplementary Information)~\cite{tsang2009centroid,Gregory2025}. This reassignment combines the two correlated detection positions into a single coordinate associated with the photon pair, thereby enhancing the signal and spatial localization of the reconstructed coincidence image. Conceptually, this is analogous to pixel-reassignment methods in classical scanning confocal microscopy~\cite{Heintzmann2013}; here, however, the reassigned coordinate is the bisector of the two detection positions, rather than the coordinate between the source and detector pinholes in classical confocal imaging.
%


\needspace{2\baselineskip}
\vspace{0.5 cm}
\noindent\textbf{\large{Optical sectioning with quantum correlations}}

\begin{figure*}
    \centering
    \includegraphics[width=\linewidth]{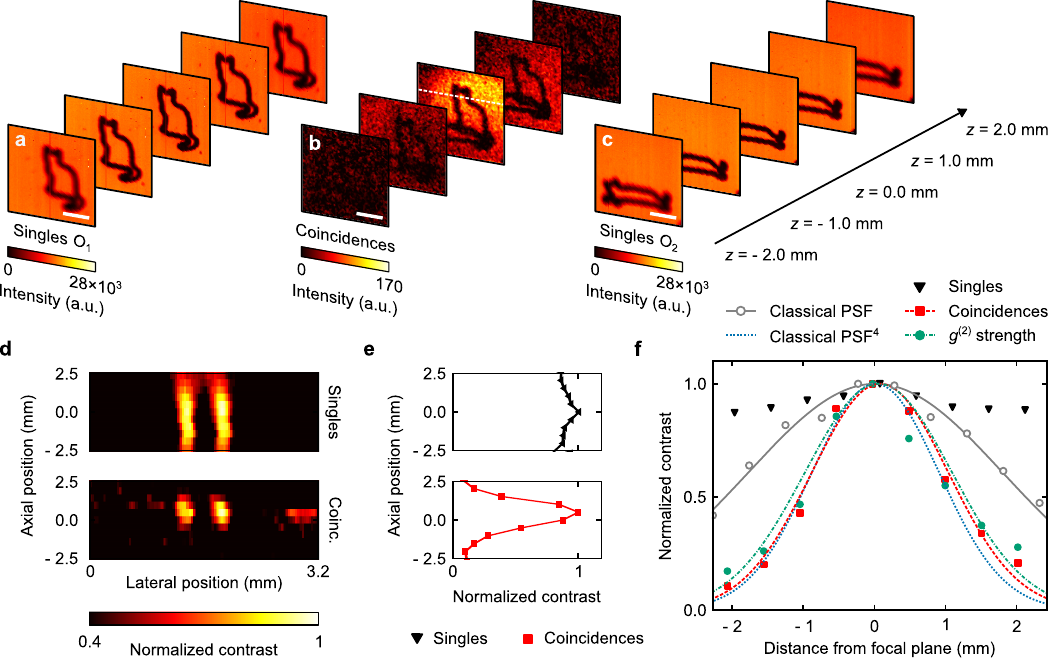}
    \caption{\textbf{Axial confinement arising from biphoton intensity correlations.} (\textbf{a}--\textbf{c})~Classical widefield images of $O_1$ (\textbf{a}), biphoton coincidence images (\textbf{b}), and classical widefield images of $O_2$ (\textbf{c}) acquired at varying axial positions $z$. Scale bars, 1 mm. (\textbf{d})~Lateral-axial ($x$-$z$) contrast maps extracted along the white dashed profile indicated in (\textbf{b}) for classical (top) and coincidence (bottom) detection. (\textbf{e})~Summed lateral intensity along $z$ for classical and coincidence imaging, showing enhanced axial confinement in the coincidence signal. (\textbf{f})~Axial response comparison. The classical widefield PSF broadens gradually with defocus whereas the coincidence signal displays a substantially reduced axial width. The classical response to the fourth power and measured $g^{(2)}$ peak amplitude are shown for comparison. Contrast data are shown as solid scattered points, PSF data is shown as hollow points, Gaussian fits are overlaid. 
    }
    \label{fig:defocus}
\end{figure*} 

\noindent To probe axial confinement arising from biphoton correlations, we translate the detection plane through the focus while recording both classical widefield (singles) and coincidence images of the dual-object configuration. To ensure high-fidelity estimation of coincidence statistics, the image from each axial plane was reconstructed from $2\times10^6$ acquired frames for both the singles and coincidence measurements. Classical widefield (singles) images of $O_1$ and $O_2$ blur progressively with defocus, as expected, whilst maintaining their overall brightness (Fig.~\ref{fig:defocus}a, c), whereas coincidence images are only visible near the focal plane (Fig.~\ref{fig:defocus}b). From the cross-section highlighted in Fig.~\ref{fig:defocus}b, we extract lateral-axial contrast maps to visualize the axial confinement of the coincidence signal (Fig.~\ref{fig:defocus}d). For visualization, transmission dips were converted to positive contrast before normalizing and lightly smoothed with a Gaussian kernel ($\sigma_x=\sigma_y=0.5$ pixels). Low-contrast background regions were also cropped to enhance the visibility of bright axial peaks. The full dynamic range images are provided in the Supplementary Information. Summing the intensity at each plane further highlights this confinement: the coincidence signal remains visible near the focus, whereas the classical signal spreads over a broader axial range (Fig.~\ref{fig:defocus}e). 

To compare axial responses quantitatively, we calculated the coincidence image contrast by taking the ratio of the masked ``cat'' regions with respect to the background (region outside the mask) - see Methods for details. When looking at the image contrast for the classical widefield case (singles), we see that the contrast remains approximately flat since the relatively large object extent averages over the underlying point-spread response -- illustrated by plotting the singles response from Fig.~\ref{fig:defocus}e on Fig.~\ref{fig:defocus}f. To provide a comparison of axial confinement, we extracted the classical widefield axial PSF from the singles measurements. This PSF was estimated from the depth-dependent blurring analysis performed along a sharp edge in the classical image (see Supplementary Information), providing an approximation of the expected diffraction-limited axial response for the measured optical system. The resulting axial profiles are shown in Fig.~\ref{fig:defocus}f. While the classical widefield PSF weakly broadens and decreases with defocus, the coincidence signal exhibits a markedly narrower axial envelope, demonstrating the optical sectioning arising from the biphoton correlations. 

The Gaussian fits to the measured axial profiles in Fig.~\ref{fig:defocus}f yield an axial width approximately $\sim 1.9\times$ smaller for the coincidence response than for widefield imaging, in direct agreement with the predicted twofold narrowing arising from coherent addition of the biphoton propagation phases. 
Moreover, the measured coincidence response closely reproduces the predicted Gaussian non-separable-biphoton response, with quantitative agreement in both axial width and profile shape compared to the fourth-power widefield PSF. The simultaneous agreement in both shape and axial localization provides compelling experimental evidence that the enhanced optical sectioning is a direct consequence of non-separable photon correlations. Unlike conventional confocal imaging, this entanglement-enabled sectioning requires neither a physical pinhole nor mechanical scanning, and demonstrates a genuine quantum advantage in axial sectioning.

%
As an independent measure of the axial response, we also compute the normalized second-order correlation function $g^{(2)}$ between the two detection regions. The fitted peak amplitude of $g^{(2)}$ follows the same axial dependence as the reconstructed coincidence signal (Fig.~\ref{fig:defocus}f), consistent with the interpretation that the confocal effect arises directly from second-order intensity correlations between the photon pairs.

\begin{figure}
    \centering
    \includegraphics[width=\linewidth]{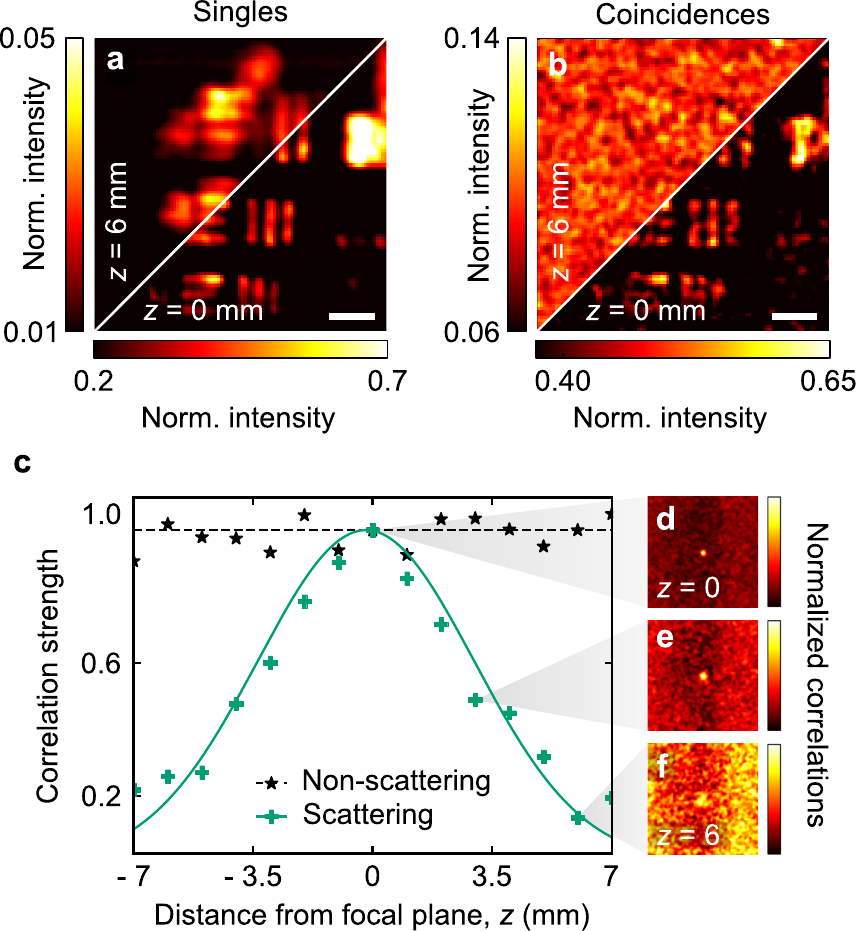}
    \caption{\textbf{Scattering-induced degradation of biphoton correlations.} (\textbf{a}, \textbf{b})~Classical singles (\textbf{a}) and coincidence (\textbf{b}) images of the scattering test object acquired at focus and under defocus. Each image is normalized to the in focus condition. Scale bars, 1~mm. (\textbf{c})~Normalized axial dependence of the excess correlation peak amplitude (relative to the background) as a function of object displacement for non-scattering (median line) and scattering (Gaussian fit) configurations. (\textbf{d}--\textbf{f})~Representative correlation peaks corresponding to three axial positions indicated in (\textbf{c}), illustrating the degradation of correlation contrast under defocus in the presence of scattering. Each image is normalized to its maximum value.}
    \label{fig:scatter}
\end{figure} 


\needspace{2\baselineskip}
\vspace{0.5 cm}
\noindent\textbf{\large{Preservation of biphoton correlations under scattering}}

\noindent To highlight the role of correlations in defining the axial sectioning, we replace the transmissive object with a weakly scattering object in the signal path, while leaving the idler path in free-space transmission and maintaining the same biphoton detection geometry (Fig.~\ref{fig:setup}). This removes the dual-object configuration used previously, reducing the system to a more standard single-object imaging geometry, without altering the underlying detection scheme. In this configuration, axial displacement is applied to the scattering object rather than the detector, allowing us to probe how the coincidence signal depends directly on the coincidence detection process under the effects of scattering. 

In the absence of scattering, the axial displacement of the object defocuses the image while preserving the underlying biphoton correlation structure, and therefore does not affect intensity of the coincidence image along the $z$ direction. Introducing a weak scatterer exactly in the image plane does not alter this behavior, as the correlations remain sufficiently well preserved. Moving the scatterer away from the image plane causes the correlations to progressively degrade. The measured signal in the coincidence image is therefore sensitive to the integrity of transverse correlations at the scattering plane and hence the axial position of a scattering object. 

To demonstrate this axial sectioning in the presence of a scatterer, Fig.~\ref{fig:scatter}a, b show representative widefield (singles) and biphoton (coincidence) images of a scattering USAF target at focus and under defocus (each reconstructed from $12\times10^6$ frames). In the singles image, defocus introduced by translating the object leads to the expected loss of spatial resolution, consistent with conventional intensity-based imaging. In the coincidence channel, image intensity is instead mediated by the survival of a measurable correlation signal, which becomes increasingly degraded under axial displacement in the presence of scattering -- resulting in confocal depth sectioning. 

Figure~\ref{fig:scatter}c shows the axial dependence of the correlation strength as an object is translated along the optical axis, comparing non-scattering and scattering conditions. In the non-scattering case, the correlation signal remains approximately stable under object translation, consistent with the preservation of transverse biphoton correlations. By contrast, the introduction of scattering leads to a rapid reduction in the measurable correlation peak with axial displacement, reflecting the progressive loss of joint detection probability. These data were acquired with $10^4$ and $10^5$ frames per axial position for the non-scattering and scattering cases, respectively, to ensure comparable signal-to-noise ratios under reduced correlation strength in the scattering regime. 

Representative coincidence maps corresponding to the axial positions highlighted in Fig.~\ref{fig:scatter}c are shown in Fig.~\ref{fig:scatter}d--f, illustrating the progressive degradation of correlation strength as the scattering object is moved away from the focal plane. Each map is normalized to its maximum for visual comparison. Taken together, these observations demonstrate that the observed axial confinement is not determined solely by optical defocus, but by the preservation of the transverse correlations. Scattering does not alter the imaging geometry but directly suppresses the measurable correlation strength, thereby restricting coincidence image formation to the axial region in which the biphoton overlap condition is satisfied. 


\vspace{0.5 cm}
\noindent\textbf{\large{Conclusions}}

\noindent Confocal microscopy conventionally achieves optical sectioning through spatial filtering of out-of-focus light. Here we demonstrate that a similar confocal effect can instead emerge inherently from biphoton coincidence detection. In this biphoton approach, axial sectioning is encoded in the spatial correlations of entangled photon pairs and recovered directly through the measurement, without physical pinholes, scanning, or engineered illumination patterns. Furthermore, the entangled state provides a quantum advantage in axial localization, yielding a $2\times$ narrowing of the axial response relative to widefield imaging, compared with the $\sqrt{2}$x narrowing of conventional confocal microscopy.

%
The equivalence between our quantum approach and traditional confocal microscopy becomes apparent when the system is interpreted using Klyshko's advanced wave picture~\cite{Klyshko1988}, in which the spatial localization of photon pairs at the pixel level corresponds to the role of the spatial pinholes in conventional confocal imaging. This picture provides an intuitive understanding of how physical pinholes can be substituted with the biphoton correlations, but does not alone capture the quantum advantage that arises from the non-separability of entangled biphoton states. 
Indeed, it is the coherence of the non-separable biphoton states that leads to a doubling of the propagation phase and that can be interpreted as a $z\rightarrow2z$ re-scaling of the  axial response. Thus, the quantum implementation removes the requirement for lateral scanning (by replacing the physical pinholes with biphoton momentum-position correlations, as captured by the Klyshko interpretation) while providing an additional phase-based mechanism for improved axial localization.

Our proof-of-principle experiments use label-free imaging, allowing the full physics of quantum confocal imaging to be observed. However, the correlation-based confocal mechanism is not intrinsically restricted to label-free samples and can be extended to fluorescence imaging. In this regime, fluorescence destroys the phase coherence associated with the transverse-wavevector correlations of the incident biphoton state while retaining spatial correlations at the detector. The phase-based enhancement of axial localization is therefore lost, while the correlation-induced spatial constraint remains, providing the conventional confocal scaling (see Eq.~\eqref{g2_sep}) while retaining the parallel widefield nature of the quantum measurement.

The frame rate of the present implementation is limited only by photon-detection throughput rather than by any fundamental aspect of the approach. Rapid advances in large-format single-photon detector arrays are expected to increase coincidence imaging rates by several orders of magnitude~\cite{Ulku2018, Wayne2022, Defienne2021}.

Because our approach is inherently parallel, increasing detector throughput could ultimately enable acquisition rates exceeding those of point-scanning confocal systems, while preserving optical sectioning without pinholes or scanning. More broadly, our results establish a direct connection between quantum correlation measurements and a foundational principle of microscopy, showing that functionality traditionally implemented through optical hardware can instead arise directly from quantum states.



\vspace{0.5 cm}
\noindent\textbf{\large{Methods}}

\noindent\textbf{Experimental setup}

\noindent The pump source was a quasi-continuous-wave Nd:YAG laser (JDSU Xcyte CY-355-150) emitting at 355 nm with an average power of 150 mW, a repetition rate of $100 \pm 10$ MHz, and a pulse duration exceeding \mbox{10 ps}. The beam was spatially filtered and expanded using a 50-mm lens, a 50-\textmu m pinhole, and a 200-mm lens to produce a high-quality Gaussian spatial mode. The pump was propagated onto a BBO crystal ($10 \times 10 \times 3$ mm) cut for type-II phase matching, generating signal and idler photon pairs at the degenerate wavelength of 710 nm with anti-correlated transverse momenta. 

Residual pump light was removed immediately after the crystal using a dichroic mirror (Chroma T4551pxt; cutoff 455 nm, 98\% transmission at 710 nm), which redirected unconverted pump photons out of the detection path. The down-converted photons were spectrally filtered using a tunable interference filter (Chroma ET710/10m; 10 nm top-hat bandpass) mounted on a tiltable stage, in combination with a fixed interference filter (Semrock FF01-711/25; 25 nm top-hat bandpass) positioned directly in front of the detector. The tilt angle of the 710-nm filter was adjusted to maximize transmission at the degenerate wavelength, resulting in circular signal and idler far-field emission patterns of equal diameter. 

The far field of the SPDC emission was imaged onto an EMCCD camera (Andor iXon Ultra 897) comprising a $512 \times 512$ pixel array with 16 \textmu m $\times$ 16 \textmu m pixel size, operated at $-90^{\circ}$C using combined internal Peltier and external water cooling. The EMCCD was operated with an electron-multiplying gain of 1,000, a vertical shift speed of 0.9 \textmu s, and an exposure time of 60 ms per frame. The mean photon occupancy per frame was maintained at $\sim 1\%$ to remain in the single-photon counting regime. A two-fold demagnification of the far-field plane was applied via a telescope comprising a 300-mm lens and a 150-mm lens to optimally match the spatial extent of the transverse momentum correlations to the detector pixel size. Single-photon detection events were identified by thresholding the EMCCD output to produce binary frames. These binary detection frames were convolved with a Gaussian kernel ($\sigma = 0.2$) to account for the finite spatial resolution of the detection system and to suppress discretization effects at the pixel level prior to correlation analysis.

\vspace{0.5 cm}
\noindent\textbf{Far-field correlation analysis}

\noindent Spatial correlations between signal and idler photons were quantified by comparing photon detection events within corresponding regions of interest in the two beams. To compensate for the intrinsic momentum anti-correlations of the SPDC emission, one region of interest was rotated by $\pi$ so that the correlated transverse momenta were mapped to the same spatial coordinates. 

For each detected photon pair, a joint detection map was constructed from the product of the Gaussian-blurred, thresholded single-photon detection events in the two regions of interest. The position of maximal joint probability was identified and reassigned to its midpoint (bisector) coordinate, corresponding to the center of mass of the correlated detection events. This bisector-based pixel reassignment concentrates correlated photon pairs into a single spatial bin, thereby improving the signal-to-noise ratio of the reconstructed correlation image. 

Accidental coincidences arising from detector dark counts and stray light were removed by subtracting a background signal obtained from temporally uncorrelated frames separated by an interval exceeding the photon coherence time. Correlation maps were computed from these background-subtracted, pixel-reassigned data, with the resulting peak height providing a quantitative measure of the photon-pair correlation strength. 

To increase the signal-to-noise ratio of the resulting quantum image, the correlation maps were subsequently convolved with a light Gaussian kernel. In the case of the scattering object, this blurred image was then enhanced using an unsharp-mask filter to improve the visibility of spatial features (this was also applied to the classical image for consistency). 


\vspace{0.5 cm}
\noindent\textbf{\large{Data availability}}

\noindent All data supporting the findings of this study will be made available on the University of Glasgow's public data repository upon publication.


\vspace{0.5 cm}
\noindent\textbf{\large{Acknowledgements}}

\noindent We thank I. Starshynov for his assistance in preparing the transmissive sample imaged in Fig.~\ref{fig:defocus}, S. P. Mekhail for his assistance in preparing the scattering sample imaged in Fig.~\ref{fig:scatter}, and K. Roberts for providing the initial camera-control software. We also thank A. Fatima and N. Westerberg for helpful discussions about the mathematical description of quantum confocal imaging. The authors acknowledge funding from the Engineering and Physical Sciences Research Council (UK, grant no. EP/Z533166/1). D.F. is supported by the Royal Academy of Engineering through the Chairs in Emerging Technologies program. M.J.P. acknowledges financial support from the Royal Society (RSRP/R1/211013P). 


\vspace{0.5 cm}
\noindent\textbf{\large{Author contributions}}

\noindent E.M.: conceptualization, methodology, software, formal analysis, investigation, visualization, writing -- original draft. E.P.: conceptualization, methodology, investigation, writing -- review \& editing. D.F.: conceptualization, supervision, project administration, funding acquisition, writing -- review \& editing. M.J.P.: conceptualization, methodology, software, resources, supervision, project administration, funding acquisition, writing -- review \& editing. 


\vspace{0.5 cm}
\noindent\textbf{\large{Competing interests}}

\noindent The University of Glasgow has filed a patent application for the technique outlined in this work in the United Kingdom (GB), application no.  2613590.5, titled ``Quantum confocal microscopy,'' on 12 June 2026, status: patent pending. All authors are named inventors on this patent application. The authors declare no further competing interests. 


\bibliography{biblio}


\end{document}